# Variations of solar electron and proton flux in magnetic cloud boundary layers and comparisons with those across the shocks and in the reconnection exhausts

Y. Wang[1,2], F. S. Wei[1], X. S. Feng[1], P. B. Zuo[1], J. P. Guo[1], X. J. Xu[3] and Z. Li[4,5]

1 SIGMA Weather Group, State Key Laboratory for Space Weather, Natinal Space Science Center, Chinese Academy of Sciences, Beijing, 100190, China; yw@spaceweather.ac.cn
2 Collegel of Earth Science, Graduate University of the Chinese Academy of Sciences, Beijing 100049, China
3 Institute of Space Science and Technology, Nanchang University, Nanchang 330031, China
4 National Key Laboratory of Science and Technology on Aerospace Flight Dynamics, Beijing, China, 100049
5 Beijing Aerospace Control Center, Beijing 100094, China

**Abstract** Magnetic cloud boundary layer (BL) is a dynamic region formed by the interaction of the magnetic cloud (MC) and the ambient solar wind. In the present study, we comparatively investigate the proton and electron mean flux variations in the BL, in the interplanetary reconnection exhaust (RE) and across the MC-driven shock by using the Wind 3DP and IMF data from 1995 to 2006. In general, the proton flux has higher increments at lower energy bands compared with the ambient solar wind. Inside the BL, the core electron flux increases quasi-isotropically and the increments decrease monotonously with energy from ~30% (at 18 eV) to ~10% (at 70 eV); the suprathermal electron flux usually increases in either parallel or antiparallel direction; the correlation coefficient of electron flux variations in parallel and antiparallel directions changes sharply from ~0.8 below 70 eV to ~0 above 70 eV. Similar results are also found for RE. However, different phenomena are found across the shock where the electron flux variations first increase and then decrease with a peak increment (>200%) near 100 eV. The correlation coefficient of electron flux variations in parallel and antiparallel directions is always around 0.8. The similar behavior of flux variations in BL and RE suggest that reconnection may commonly occur in BL. Our work also implies that the strong energy dependence and direction selectivity of electron flux variations, which are previously thought to have no enough relevance to magnetic reconnection, could be considered as an important signature of solar wind reconnection in the statistical point of view.

# 1 Introduction

Magnetic clouds (MCs) are large-scale transient structures in the solar wind. In the past few decades, problems about their solar origin, magnetic field and plasma structures have been widely investigated (Bothmer & Schwenn 1994; Burlaga et al. 1980; Burlaga et al. 1981; Farrugia et al. 1994; Lepping et al. 2006; Lepping et al. 1997). In addition, as a subset of the Interplanetary Coronal Mass Ejections (ICMEs), the propagation of MCs in interplanetary space is also an important issue in heliospheric physics research. For example, a MC could be overtaken by a corotating stream that would compress the plasma and field and make its tail region turbulent (Lepping, et al. 1997). There might also be magnetic holes, directional discontinuities or reconnection layers in the front boundary of the MCs (Janoo et al. 1998). Therefore, the interaction between the MC body and the ambient solar wind seems to be a complex problem which not only aggravates the difficulty to understand the evolution of ICME but also increases the complexity to identify the MC boundary (Wei et al. 2006; Wei et al. 2003b).

Up to now, there is still no consistency among the criteria to identify the MC boundary such as the temperature decrease, density decrease, directional discontinuity, magnetic hole and bidirectional streaming of suprathermal electrons, as pointed out by many researchers (Burlaga et al. 1990; Fainberg et al. 1996; Farrugia et al. 2001; Osherovich et al. 1993; Wei, et al. 2006; Wei et al. 2003a; Wei, et al. 2003b; Wei et al. 2003c). Wei et al. (2003b) statistically analyzed the boundary physical states of 80 MCs detected from 1969 to 2001 and suggested that the MC boundary is a complex boundary layer (BL) formed by the interactions between the MC and the background solar wind, rather than a simple discontinuity. The BL ahead of MC is called the front BL, while the following one is the tail BL (Wei, et al. 2003b). For each BL, its outer boundary (M) is usually identified by the magnetic field intensity drop, the abrupt change of field angle, and is accompanied with the "three-high state" in the plasma beta value, temperature and density, while the inner boundary (G), which separates the interaction region from the MC body, is usually associated with the "three-low state" also in the plasma beta value, temperature and density (Wei, et al. 2003b). The MC detected by Wind on 15 May 1997 provides a typical sample to reveal the BL's properties (see Figure 1). The spacecraft at 1 AU observed in sequence the MC-driven shock (if there is), the front BL and MC body. As seen in the Figure, the magnetic field, plasma temperature and density behaviors inside the BL, which is separated by the obvious boundaries (labeled by $M_f$ and $G_f$, the subscript 'f' means front), are completely different from those in the nearby upstream solar wind, the following MC body and the preceding shock (sheath) region. Previous analyses show that the BL is often a unique structure exhibiting decreased magnetic field as well as heated and accelerated plasma. These features are preliminarily interpreted to be associated with the magnetic reconnection process since they are important manifestations that could be often observed in a magnetic reconnection region (Wei, et al. 2006; Wei, et al. 2003c).

Magnetic reconnection is an important process that can convert magnetic energy into thermal and kinetic energy. Many researchers have intensively studied its dynamics in geo-magnetosphere and solar corona, but the magnetic reconnection phenomena in the solar wind has drawn relatively less attention so far. Early study suggested that interactions between a fast ICME and the ambient solar wind might cause reconnection at the compressed leading boundary region of the ICME (McComas et al. 1994). Recently, observations of reconnections at both leading and trailing boundaries of interplanetary small-scale magnetic flux ropes were also reported (Tian et al. 2010). Previous work seems to suggest that such type of RE is more often observed in low-beta solar wind or in the interiors of ICMEs but not particularly prevalent in the leading edge of an ICME since the roughly Alfvénic accelerated flows within field reversal regions, which are regarded as the 'direct evidence' of magnetic reconnection, are hard to identify in the front region of the ICME (Gosling 2011; Gosling et al. 2005b). However, both numerical simulations and physical models have demonstrated that reconnections could occur in the front BL where MC interacts with the ambient solar wind (Dasso et al. 2006; Wang et al. 2010; Wei, et al. 2003a; Wei, et al. 2003b). Therefore, it is worthwhile to make clear that what the dominant physical process is inside the BL and whether the reconnection process plays an important role.

The magnetic field is highly related to the local plasma distribution for which the density and temperature are macroscopic manifestations of the plasma velocity distribution function (VDF). Hence, investigation on the VDF is an effective way to diagnose the magnetic field and plasma structure in the solar wind (Gosling et al. 1987; Gosling et al. 2005c; Larson et al. 1997). Generally, the solar wind electron contains the thermal core electron and the suprathermal electron with a breakpoint near ~70 eV. Electron at lower energy bands plays an important role in the calculations of electron density and temperature because they can be calculated from the zero- and second-order moments of VDF. The suprathermal electron usually contains two components, a nearly isotropic electron called halo and an electron beam coming directly outward from the Sun called Strahl. The suprathermal electron, especially Strahl electron carrying the heat flux outward from the Sun, has been widely used to diagnose the magnetic field configuration in the solar wind (Gosling et al. 1987;

Gosling et al. 2005c; Larson et al. 1997). Electron heat flux dropout in the solar wind is speculated to be an evidence for interplanetary magnetic reconnection (McComas et al. 1989). However, hardly any work has been done to demonstrate whether electron flux variations could be regarded as a sufficient signature for solar wind magnetic reconnection. The enhanced flux of energetic particles, especially the energetic electrons (>100 keV), might also indicate the existence of acceleration processes, such as magnetic reconnection (Lin & Hudson 1971; Oieroset et al. 2002; Wang, et al. 2010) or shock (Potter 1981; Tsurutani & Lin 1985). In this paper, we use the Wind data to statistically analyze the proton and electron flux variations in the BL and compare them with those in the MC-driven shock and interplanetary RE and try to reveal the dynamic process inside the BL.

## 2 Data set description and events selection

The WIND 3-D plasma and energetic particle instrument (3DP) provides full three-dimensional distribution of electrons and protons covering a wide range of (time varying) energy bands (Lin et al. 1995). The data provided by the electron electrostatic analyzers (EESA), the proton electrostatic analyzers (PESA) and the semiconductor telescopes (SST) will be analyzed. We only analyze the proton flux in the omni direction since the directional proton flux data is not available online. For the electron flux data, we investigate the flux in parallel, perpendicular and antiparallel directions. Since the measurements of electron density and the electron flux at lower energy bands are greatly affected by the instrumental restrictions, we assume that the electron density is equal to the proton density and the electron flux data below 18 eV are not used (see the discussion section). Moreover, in order to facilitate the statistical work, we reconstruct the energy bands at fixed energy to all events; and to avoid the frequently occurred invalid data, we do not use the data provided by the EESA-H and PESA-L experiments. Finally, the electron flux from 18 eV to 500 keV (EESA-L: 18, 27, 42, 65, 103, 165, 265, 427, 689 eV; SST-F: 27, 40, 66, 108, 183, 307, 512 keV) and the proton flux from 4 keV to 4 MeV (PESA-H: 4, 6, 9, 11, 15, 21, 28 keV; SST-MO: 74, 128, 197, 333, 552, 1018, 2074, 4440 keV) will be analyzed.

The BL events are identified according to the BL concept and identification criteria (Wei, et al. 2003b). The physical characteristics of the tail BL are quite different from the front BL (Wei, et al. 2003b; Wei, et al. 2003c; Zuo et al. 2007), and this paper only focus on the flux variations in the front BL detected from 1995 to 2006 (41 events are listed in Table 1a). The interplanetary RE events are chosen from the list provided by Huttunen et al. (2007). Since the time resolution of the EESA and PESA is ~98s, the RE events with too short duration (<98s) are excluded (24 events are listed in Table 1c). The MC-driven shock events are selected based on the work of Feng et al. (2011). We use the following criteria to select shock events as 'MC-driven' events (23 events are listed in Table 1c): (1) the angle $\theta$ between the axe of the MC, adopted by fitting the constant α force-free model to the magnetic fields (Feng et al. 2010), and its leading shock normal is in the range from 65 to 115 degree; (2) the interval between the shock and the beginning of the MC is less than 14 hours.

During the statistical work, we quantify the flux variations in the form of $\Delta F=(F_2-F_1)/F_1$ at each energy band and direction for each event. In the case of the BL events, $F_2$ is the mean flux inside the BL and $F_1$ is the mean flux of the nearby upstream solar wind with 30 minutes duration. For RE, $F_2$ is the mean flux inside the RE and $F_1$ is mean flux of the nearby upstream solar wind with the same duration of the RE. For the shock, $F_1$ and $F_2$ are the mean flux of upstream and downstream solar wind respectively with 12 minutes duration and 3 minutes away from the shock discontinuity. The possible influences of our selection criteria and sample method on the final results of flux variations will be discussed in the last section.

## 3 Statistical results

The local magnetic field and plasma parameters of the three types of events BLs, REs, Shocks) are listed in Table 1. It is found that the magnetic field decreases ($\Delta Bt \sim -16.4\%$) in most of the BL events and plasma is usually compressed ($\Delta Np \sim 42.9\%$) and heated ($\Delta Te \sim 5.3\%$, $\Delta Tp \sim 16.6\%$). These phenomena resemble previous work (Wei, et al. 2003b; Wei, et al. 2006) on BL events, and they are quite similar to the RE events despite the somewhat larger temperature increment ($\Delta Bt$, $\Delta Np$, $\Delta Te$, $\Delta Tp \sim -20.1\%$, 35.8%, 10.6% and 27.1% respectively). The average duration of the BLs ($\Delta t \sim 67$ minutes) is ~18 times longer than that of the REs ($\Delta t \sim 229$ seconds) and the absolute difference of proton velocity in the REs ($\Delta Vp \sim 21.7$km/s) is larger than that in the BLs ($\Delta Vp \sim 12.1$km/s). The MC-driven shocks are usually fast forward shocks across which the magnetic field, proton and electron temperature and plasma speed always increase much (the average changes of $\Delta Bt$, $\Delta Np$, $\Delta Te$, $\Delta Tp$, $\Delta Vp$ are ~140.7%, 122.2%, 82.9%, 161.3% and 81.5km/s respectively) . It is also noted that there are few strong strength MC-driven shocks. The obtained density compression ratio of the shock is in the range of 1.3-4.6 with a mean value of only 2.2.

Figure 2 presents the electron flux variation $\Delta F$ averaged over all events in the parallel, anti-parallel, perpendicular direction as well as the omni proton flux variation also averaged over all events. The flux variations (flux decrease or increase) depend both on the direction and the energy. Inside the BL, the core electron flux in the parallel, anti-parallel and perpendicular direction increase consistently and the increment amplitude decreases with energy monotonously from ~30% (at 18eV) to ~10% (at 70eV); the increments of suprathermal electron (100-700eV) in the parallel and antiparallel directions are very small (<4%), but it is noted that their standard error are obviously large; the energetic electron (>100keV) also has slight increments in the perpendicular direction; the increments of the proton omni flux fall at higher energy bands but they have a prominence around 70keV. In the RE, although the energetic electron in the parallel direction has higher increment with larger standard error, the flux variations have similar behaviors compared with the BL as a whole. By contrast, across the shock, flux behaviors are quite different. The electron flux variations have peak increments (>200%) around ~100eV and decline on both sides; we also note that they have higher increments in the perpendicular direction and the corresponding energy of the peak increment is also higher in the perpendicular direction (~165eV) than in the field-aligned direction (~65eV); the omni proton flux increments decrease monotonously from ~280% (at 4keV) to ~10% (at 4MeV).

During the statistical work, it is also found that the correlations of the electron flux variations in parallel and antiparallel directions have a sharp change around 70eV where solar wind magnetic reconnection occurs. Figure 3 provides the correlation coefficients of electron flux variations in the parallel and antiparallel directions. In all events, the core electron has (strong) positive correlations (BL and RE: r>0.8; shock: r>0.6); while the suprathermal electron in the BL and RE has very low or negative correlations (BL: r~0; RE: r~-0.2), in addition, the correlations are even lower across (downstream to upstream) the RE (r~-0.4); however, no obvious changes are found across the shock which always has high correlation around 0.8.

## 4 Explanations for flux variations

Since the compressing and heating effects are quite common inside the BL, RE and across the shock, these effects could account for the presented core electron flux variations. The zero-order moment of the VDF is equal to the mass density. Accordingly, if the VDF has a Maxwellian distribution, the density will behave essentially the same as the flux. Especially, the lower the flux energy is, the more similar behaviors the density and flux have. As is seen in Figure 4, the isotropic increments of electron flux at 15-41eV vary consistently with the density changes in the BL. The final increments of the electron flux at 18eV are also roughly consistent with the average density increments (listed in Table 1a) and previously statistical results (Wei, et al. 2006). Therefore, the enhancements of the core electron flux with high correlation in all three directions could be related to the density increase in the compressed BLs. The core electron flux variation in the REs is similar to the BLs, however, it behaves totally different across the MC-driven shocks. The electron flux increments ascend first and then descend with peak value near ~65 eV and ~165 eV in the field-aligned and perpendicular directions respectively. Such flux behaviors could not be merely caused by the density increase and increments are also inconsistent with the average density increments. We noted that the increase of electron temperature across the shock is quite higher than that in the BL and RE. Since the 'moment temperature' (Burlaga 1995) is calculated from the second-order moment of the VDF, we speculate that the increments of electron flux with hill-like shape are mainly dominated by the heating effect of the shock. This result is consistent with previous observations which show that the inflated electron VDF caused by heating in both the parallel and perpendicular directions is always found downstream of the shock (Fitzenreiter et al. 2003). In addition, according to early researches, for weaker shocks, the electron heating was primarily perpendicular to the magnetic field due to the conservation of magnetic moment (Feldman et al. 1983). The present statistical results with higher flux increments in the perpendicular directions could be also supported by such explanations, since many of our selected MC-driven shocks have relatively small density compression ratio.

As described in the introduction section and the references therein, the bidirectional, unidirectional and absence of Strahl electron could reflect the configurations of closed, open and disconnected magnetic field lines from the Sun respectively (Gosling, et al. 1987; Gosling, et al. 2005c; Larson, et al. 1997). Although previous work has speculated their dependences and analyzed their behaviors in the reconnection (Gosling, et al. 2005c; McComas, et al. 1989), there are no sufficient direct relevance established between the electron flux variations and solar wind magnetic reconnection. In our statistical work, we find that the suprathermal electron (100-700eV) flux displays low or even negative correlation between the parallel and antiparallel directions when a spacecraft across the RE. Here we would like to explain why these features are related to the solar wind reconnection in some details. As sketched in Figure 5, taking the Strahl electron in ideal case for instance, the intensity of flux is simply normalized by only two arbitrary quantities: 100 (obvious Strahl electron) and 10 (no obvious Strahl electron). The flux status is described by [F0,F180], where F0 and F180 stand for the flux of Strahl electron in the parallel and antiparallel directions respectively. Accordingly, the status of bidirectional Strahl electron, unidirectional Strahl electron in the parallel and antiparallel directions and no obvious Strahl electron could be described by [100,100], [100,10], [10,100] and [10,10] respectively. In cases I, the spacecraft would detect decreased and unchanged Strahl electron in the parallel and antiparallel directions respectively inside the RE, and the increments are [-90,0] ([10,10]-[100,10]). Similarly, the increments in case II, III and IV are [0,90], [0,-90] and [90,0] respectively. Therefore, in statistical analyses, the correlations of the Strahl electron flux variations in parallel and antiparallel directions should be low if the spacecraft

231  encounter the above four cases randomly. Mathematically speaking, both the correlation coefficient

232  and the averaged increments should approach 0. Moreover, across the RE, the increments become

233  [-90,90] in case I, II and [90,-90] case III, IV. We could see that they always reveal anti-correlated

234  relations in the parallel and antiparallel directions. Accordingly, the theoretically computed

235  correlation coefficient is even lower (should be -1) in the statistical work. Certainly, our assumptions

236  are relatively simple, for example, the real flux intensity could not be only two quantities (100 and

237  10), and thus the finally obtained correlation coefficients and mean flux increments might not as

238  ideal as in the analyses. However, the flux variations of suprathermal electron still reveal the

239  properties that the mean increments approach 0 with large stand errors and the correlation

240  coefficients are low (~-0.2) and lower (~-0.4) in and across the REs. Other effects, such as particle

241  scattering, could also modify the flux of electron. If so, it should be explained why the correlation of

242  core electron are always higher than the suprathermal electron and why the correlation coefficients

243  change sharply around ~70eV. Perhaps the correlation coefficients should change more smoothly if

244  the scattering process plays a dominant role. In addition, since these RE events are not magnetically

245  connected to the Earth's bow shock (Huttunen et al. 2007), the obtained results would not be greatly

246  affected by particle reflection either. Moreover, the correlation coefficients across the MC-driven

247  shock, in which there is no obvious break or reverse of magnetic field line, are always high

248  (~0.7-0.9). For these reasons, we tend to regard that the solar wind magnetic reconnection is the best

249  candidate process that could account for the statistically obtained low or negative correlations of

250  suprathermal electron increments in the parallel and anti-parallel directions.

251

252  The energetic electron might come from the Sun directly just as the suprathermal electron, but it should be noted that its flux variations (mean value and stand error) are somewhat different compared with the suprathermal electron. Shocks could also accelerate electron to high energy, but early work (Potter 1981; Tsurutani & Lin 1985) shows that the shock associated accelerations of energetic electron are weak near 1AU. We also find that the energetic electron sometimes shows a pike or step-like increase across the shock with short durations, thus the increments are smoothed by the sample method (3 minutes away and 12 minutes average). These reasons might be responsible for the relatively slight flux increments of energetic electron across MC-driven shock in our statistical work. Previous work (Gosling et al. 2005a) suggests the absence of energetic electron inside the RE. However, observations (Lin & Hudson 1971; Oieroset, et al. 2002) show that energetic electron produced by magnetic reconnection does exist, and the energetic electron could also be found inside the BL (Wang, et al. 2010). As pointed out by Wang, et al. (2010), the MC driving reconnection would prefer to generate complex structures (e.g. diverse magnetic islands) rather than form a single X line in the reconnection region under real solar wind conditions. These seemingly irregular structures could probably play an important role in the generation of energetic electrons (Ambrosiano et al. 1988; Drake et al. 2006; Goldstein, Matthaeus, & Ambrosiano 1986; Matthaeus, Ambrosiano, & Goldstein 1984; Wang, et al. 2010). Actually, for 6 out of 24 RE events and 6 out of 41 BL events, the omni flux increments of ~512 keV electrons exceed 30%. Therefore, the energetic electron might be the shock associated or come from the Sun directly, and it could also be generated by reconnection. However, as analyzed above, the energetic electron accelerated by reconnection seems to be a more reasonable cause for the flux increments inside the REs and BLs.

## 5 Discussion and summary

The data provided by the PESA-L and PESA-H detectors has many gaps and the energy bands are time varying. Due to the lack of reliable and detailed data, it is hard to explain proton flux variations comprehensively or draw a conclusion definitely by only using the omni directional data above 4keV. We still note that the final statistical result of proton in the BL is similar to the previous single-event observation (Wang, et al. 2010) in which a flux peak around 70keV is also found. Since the proton flux increment (~280%) across the MC-driven shock at 4keV is higher than both the density and temperature increments (~122% and 161%, respectively), it is speculated that the proton VDF around 4keV might also be inflated as that of the electron near ~70eV across the shock.

The electron flux data at low energy bands should be calibrated before use. However, the technical calibrations such as assuming a Gaussian fitted VDF and estimating the spacecraft potential are troublesome. The accuracy is not well guaranteed either. So we do not process the electron flux data below 18 eV. Actually, the electron flux data at these low energy bands seems not to affect the main conclusions of the paper, and neither does the electron flux data provided by EESA-H (1-20 keV). Besides, the electron density in the solar wind could also be calibrated by searching a bright 'plasma line' in the frequency spectrogram of the wave. It is found that the difference between the proton density and the calibrated electron density in our statistical work is very small, so the proton density is used to replace the electron density.

As mentioned above, the presented sample method by choosing the mean flux in the specified time range might have some smooth effect on the flux variations. Actually, we also adopted other sample method, such as applying the maximum flux in the same time range to all the events.

Although we got more unsmooth results, the main features also resemble the results presented here. We also try to change the criteria of the 'MC-driving' shocks, and we find that the main conclusions of the paper are not changed either, despite that the angle between the axe of the MC and shock normal and the interval between the shock and the beginning of the MC would affect the final increments to some extent.

Magnetic field decrease, density and temperature increase are similar in the RE and BL, and similar flux variation behaviors are found between these two structures. Hence we suggest that the flux variations in the BL are mainly related to the magnetic reconnection process. However, as preliminarily discussed in the introduction section, some researchers pointed out that the roughly Alfvénic accelerated plasma flows--the 'direct evidence' (Gosling 2011; Gosling, et al. 2005b), are rarely identified inside the front BL (except two events: 20001003 and 20040724). At first, it should be recognized that no roughly Alfvénic accelerated plasma flows does not mean no magnetic reconnection, since the reconnection jets might not be measured or the generated jets do not meet the referred criteria. Previous simulations (Wang, et al. 2010; Wei, et al. 2006; Wei, et al. 2003a) imply that the BL has strong turbulent property under high magnetic Reynolds number condition ($R_m \sim 10^4$). While, as also suggested by Matthaeus et al. (2003), turbulence should commonly drive reconnection in the solar wind. Inside the BL, the compression of the MC behaves as driving flows that would reduce the characteristic thickness of the local current sheet from $\sim 10^8$ km (in the corona) to $\sim 10^3$ km (in the solar wind). Accordingly, the magnetic Reynolds number could decrease from $\sim 10^{10}$ to $\sim 10^4$. Besides, the magnetic field inside the BL always shows abrupt deflections in the field direction. If the frozen field theorem is locally broken, these conditions are all favored by the potential magnetic reconnection (Wei, et al. 2006). Actually, in many cases, the BL is a complex layer with turbulent and irregular structures, besides, the trajectory of the spacecraft relative to the orientation of RE is not always suitable for the observation. So the roughly Alfvénic accelerated plasma flows that completely meets the reconnection criteria as those reported events might be hard to identify (Gosling, et al. 2005b; Huttunen, et al. 2007; Phan et al. 2006; Tian et al. 2010; Wang, et al. 2010; Xu, Wei, & Feng 2011). In addition, the referred criteria, especially the jets, (Gosling 2011; Gosling, et al. 2005b; Paschmann et al. 1986; Sonnerup & Cahill 1967) are described as 'a useful guide' (Sonnerup et al. 1981) for the identification of reconnection and have made wonderful achievements in the realm of magnetic reconnection, yet it should still be cautious to use such criteria because they are obtained under the MHD descriptions with the assumption of ideal reconnection model. Remarkably, it is pointed out (Sonnerup, et al. 1981) that such criteria have never been demonstrated to be 'incontrovertible'. Recent simulations also show that the outflowing reconnection jets could even turn back and link with the inflows to form closed-circulation patterns in turbulent reconnection (Lapenta 2008). Accordingly, reconnection generated plasma flows might not meet the referred criteria strictly in real three-dimensional space. Therefore, it is quite possible that many reconnections inside the BL do occur and the reconnection jets are indeed measured. However, they are excluded by the criteria so that many researchers think there is no reconnection. We do not want to discuss the reconnection criteria further since it is beyond the scope of this paper. Other factors should also be taken into consideration carefully, such as the life span and the evolution of the reconnection itself. As studied previously (Wei, et al. 2003a; Wei, et al. 2003b), the magnetic reconnection might not be ongoing process all the time. After the reconnection occurs, the reconnection conditions would be weakened and the frozen-in condition would be gradually recovered until the local condition is ready for the next potential magnetic reconnection. Since this process might continue to repeat itself, a single spacecraft across the BL might observe the 'remains' or the 'preorder' of magnetic reconnection. For these reasons, the signatures of reconnection, such as the Alfvénic accelerated flows (Gosling, et al. 2005b; Huttunen, et al. 2007; Phan et al. 2006; Tian et al. 2010; Wang, et al. 2010; Xu, Wei, & Feng 2011), might be not prominent to be identified sometimes. However, we have reasons to believe that the electron flux variations would not be affected and could reflect the field topological structure of the magnetic reconnection event to a certain extent.

In summary, we carry out a statistical study to analyze the proton and electron flux variations inside BL events on reliable energy bands and compare them with those in the RE and across the MC-driven shocks. The results show that the BL is a unique complicated transition layers that displays some reconnection characteristics. The core electron flux behaviors inside the BL and RE are related to the density increase. The hill-like electron flux increments across the shock are mainly dominated by the temperature increase. It is also found that the correlations of the electron flux variations in parallel and antiparallel directions have a sharp change around ~70eV where solar wind magnetic reconnection occurs. The correlation coefficients of the suprathermal electron in the parallel and antiparallel directions are found to be low. Further analyses imply that strong energy dependence and direction selectivity of flux variations could be regarded as an important signature of solar wind reconnection in the statistical point of view.

## Acknowledgement

This work is jointly supported by the National Natural Science Foundation of China (40890162, 40904049, 40921063, and 41031066), 973 program 2012CB825601 and the Specialized Research Fund for State Key Laboratories. We thank NASA CDAWEB and Space Sciences Laboratory at UC Berkeley for providing the Wind 3DP and MFI data. Y. Wang thanks T. R. Sun, H. Tian and H.Q. Feng for their helpful discussions.

**Figure captions**

Figure 1 Wind measurements of the magnetic field strength, latitude angle, azimuth angle, proton density, temperature, velocity, pressure and plasma beta value on 1997 May 15. The shock is indicated by the dotted lines; the dashed lines marked by $M_f$ and $G_f$ represent the BL region; the following MC body is also indicated.

Figure 2 Normalized mean flux variations with error bars at each energy band. First row: BL; second row: RE; third row: MC-driven shock; first column: electron flux in the parallel direction; second column: electron flux in the perpendicular direction; third column: electron flux in the antiparallel direction; fourth column: proton flux in the omni direction.

Figure 3 Correlation coefficients of electron increments in the parallel and anti-parallel directions. In the BL: red, in the RE: green, across the MC-driven shock: blue, across the RE: black.

Figure 4 Plasma and magnetic field conditions near a BL on 05/27/1996 (marked by $M_f$ and $G_f$, dashed lines). From top to bottom: the magnitude of magnetic field, electron density, electron flux in parallel, perpendicular and antiparallel directions. The color lines represent different energy bands at 15 eV (black), 21 eV (blue), 29 eV (yellow) and 41 eV (red) respectively.

Figure 5 Schematic plot of magnetic field disconnection (possible reconnection) of 4 cases and the possibly following MC.

Table 1a. Typical magnetic cloud boundary layers observed by WIND

| No.[a] | DATE[b] | Start[c] | Dur[d] | ΔBt[e] | |ΔVp|[f] | ΔNp[g] | ΔTe[h] | ΔTp[i] |
|---|---|---|---|---|---|---|---|---|
| 1 | 19950208 | 0252 | 31 | -8.09 | 6.58 | 17.31 | -4.57 | -4.28 |
| 2 | 19950403 | 0629 | 75 | -12.91 | 1.70 | 40.91 | 0.22 | 19.85 |
| 3 | 19950822 | 2036 | 61 | -3.76 | 1.53 | 10.24 | -2.40 | 5.61 |
| 4 | 19951018 | 1820 | 41 | -22.24 | 2.38 | 4.15 | -0.62 | 1.94 |
| 5 | 19960527 | 1210 | 152 | -21.89 | 2.09 | 93.15 | -3.45 | 7.16 |
| 6 | 19960701 | 1546 | 100 | -17.55 | 9.92 | 4.22 | -1.03 | -0.47 |
| 7 | 19970411 | 0524 | 30 | 8.21 | 0.36 | -8.47 | -2.69 | 21.94 |
| 8 | 19970421 | 1152 | 13 | -30.56 | 2.36 | 7.40 | -2.37 | 12.08 |
| 9 | 19970515 | 0732 | 139 | -16.73 | 32.64 | -22.53 | 18.07 | 154.98 |
| 10 | 19970715 | 0844 | 21 | -36.69 | 3.50 | 88.51 | 0.16 | 11.94 |
| 11 | 19970803 | 1005 | 226 | -4.81 | 12.72 | 133.46 | -16.20 | -1.53 |
| 12 | 19970918 | 0255 | 57 | -18.28 | 7.70 | 40.77 | 2.92 | 17.14 |
| 13 | 19971107 | 1438 | 59 | -1.20 | 0.45 | 21.52 | -7.61 | -0.97 |
| 14 | 19971122 | 1448 | 22 | -15.12 | 23.70 | 49.51 | 9.73 | 53.52 |
| 15 | 19980502 | 1233 | 21 | -4.18 | 5.43 | 30.12 | 1.24 | 3.71 |
| 16 | 19980624 | 1611 | 31 | -18.36 | 7.02 | 40.40 | -3.93 | -9.90 |
| 17 | 19980820 | 0450 | 263 | -34.37 | 35.43 | 104.60 | -6.23 | 21.38 |
| 18 | 19981108 | 2250 | 79 | -11.65 | 5.32 | 39.97 | 17.13 | 23.54 |
| 19 | 19990218 | 1149 | 33 | -23.58 | 23.58 | 41.22 | 26.70 | -2.38 |
| 20 | 19990809 | 0756 | 142 | -7.00 | 3.47 | 30.88 | 18.78 | -6.21 |
| 21 | 20000220 | 0155 | 193 | -39.13 | 4.08 | 55.08 | 17.96 | -27.99 |
| 22 | 20001003 | 1634 | 44 | -8.62 | 19.14 | 5.86 | 17.81 | 55.14 |
| 23 | 20010421 | 2347 | 25 | -13.83 | 1.40 | 30.07 | 10.91 | 11.76 |
| 24 | 20010710 | 1638 | 92 | -5.55 | 4.90 | -0.30 | 2.92 | 15.44 |
| 25 | 20020319 | 2127 | 131 | -18.96 | 5.86 | 2.80 | 4.54 | 42.80 |
| 26 | 20020324 | 0305 | 14 | -21.83 | 4.72 | 123.21 | 23.79 | 32.09 |
| 27 | 20020418 | 0419 | 20 | -19.89 | 17.98 | 17.93 | 4.49 | 61.71 |
| 28 | 20020519 | 0246 | 34 | -33.36 | 14.79 | 43.60 | 12.33 | 5.87 |
| 29 | 20020801 | 1119 | 26 | -22.67 | 9.31 | 48.73 | 25.64 | 23.08 |
| 30 | 20020802 | 0604 | 71 | -6.07 | 1.46 | 21.70 | 1.64 | 18.36 |
| 31 | 20020903 | 0250 | 71 | 4.21 | 9.44 | 33.34 | 18.89 | -7.48 |
| 32 | 20040404 | 0205 | 18 | -18.46 | 33.65 | 191.12 | 1.19 | 62.40 |
| 33 | 20040722 | 1258 | 56 | -28.26 | 27.65 | 36.16 | 15.41 | -14.50 |
| 34 | 20040724 | 1129 | 27 | -9.27 | 12.78 | -10.59 | -5.54 | 10.99 |
| 35 | 20040829 | 1830 | 28 | -18.18 | 33.78 | 0.17 | -1.02 | 0.54 |
| 36 | 20041109 | 1937 | 53 | -41.59 | 5.98 | 68.43 | 20.51 | 13.83 |
| 37 | 20050520 | 0604 | 42 | -17.01 | 2.46 | 29.97 | -7.81 | 14.32 |
| 38 | 20050612 | 1441 | 21 | -44.36 | 81.13 | 54.60 | -2.16 | 10.77 |
| 39 | 20051231 | 1233 | 76 | -37.93 | 6.83 | 187.31 | 16.98 | 7.20 |
| 40 | 20060205 | 1759 | 63 | 3.36 | 0.57 | 5.47 | -10.34 | -13.15 |
| 41 | 20060413 | 2023 | 41 | -3.65 | 8.08 | 46.47 | 5.87 | 29.96 |
| | AVERAGE | | 67 | -16.39 | 12.05 | 42.89 | 5.31 | 16.64 |

452 Table 1b. Selected solar wind reconnection exhausts observed by WIND

| No.[a] | DATE[b] | Start[c] | Dur[d] | ΔBt[e] | |ΔVp|[f] | ΔNp[g] | ΔTe[h] | ΔTp[i] |
|---|---|---|---|---|---|---|---|---|
| 1 | 19971116 | 164250 | 220 | -38.96 | 11.13 | 157.35 | 30.76 | 96.31 |
| 2 | 19980416 | 005434 | 198 | 2.27 | 17.01 | -26.45 | 2.55 | 96.76 |
| 3 | 19980821 | 202036 | 240 | -22.31 | 9.24 | 37.94 | 16.54 | 82.14 |
| 4 | 19980917 | 033315 | 109 | -14.52 | 16.27 | 28.77 | -0.15 | 29.06 |
| 5 | 19990218 | 102624 | 218 | -23.31 | 56.20 | 213.91 | 63.95 | 7.90 |
| 6 | 19990615 | 143235 | 108 | -13.32 | 16.82 | 42.82 | 13.23 | 91.98 |
| 7 | 19990626 | 054600 | 550 | -25.24 | 7.95 | 29.85 | 20.53 | 2.85 |
| 8 | 19990728 | 043559 | 189 | -24.44 | 6.29 | 39.33 | 1.72 | 49.62 |
| 9 | 19990810 | 183820 | 356 | -43.70 | 2.26 | 29.98 | 15.48 | 16.78 |
| 10 | 19990919 | 091004 | 266 | -30.09 | 20.71 | 24.34 | 1.90 | 5.43 |
| 11 | 20000419 | 035916 | 194 | -39.35 | 18.40 | 15.77 | 10.10 | 13.30 |
| 12 | 20010617 | 163023 | 157 | -19.27 | 38.24 | 48.55 | 16.36 | 1.74 |
| 13 | 20020202 | 035725 | 260 | -32.48 | 49.74 | 65.86 | 28.72 | 54.89 |
| 14 | 20020419 | 004130 | 300 | -9.55 | 36.29 | -14.35 | -3.54 | 10.59 |
| 15 | 20020628 | 152632 | 333 | -9.36 | 14.23 | 22.16 | 5.92 | -11.21 |
| 16 | 20030302 | 210955 | 107 | -32.84 | 11.52 | 6.96 | 4.90 | 27.66 |
| 17 | 20040724 | 115110 | 235 | 7.41 | 62.34 | 5.42 | 0.87 | 45.34 |
| 18 | 20040826 | 092250 | 175 | -12.69 | 11.46 | -1.11 | -3.96 | 32.22 |
| 19 | 20040914 | 212651 | 121 | -20.80 | 60.91 | 36.30 | 12.67 | -15.88 |
| 20 | 20040919 | 064100 | 670 | -4.55 | 12.89 | 76.63 | 9.34 | 5.24 |
| 21 | 20041008 | 070545 | 130 | -3.13 | 13.19 | 13.88 | -1.56 | 6.86 |
| 22 | 20041011 | 152342 | 134 | -18.82 | 16.04 | -3.39 | 4.06 | -2.68 |
| 23 | 20041029 | 024531 | 119 | -38.80 | 9.63 | 0.77 | 0.98 | 1.80 |
| 24 | 20041206 | 022056 | 115 | -14.50 | 0.16 | 2.70 | 1.95 | 2.33 |
| | AVERAGE | | 229 | -20.09 | 21.65 | 35.58 | 10.56 | 27.13 |

453
454
455

Table 1c. Selected leading shocks ahead of magnetic cloud observed by WIND

| No.[a] | DATE[b] | Start[c] | $\Delta Bt$[e] | $|\Delta Vp|$[f] | $\Delta Np$[g] | $\Delta Te$[h] | $\Delta Tp$[i] |
|---|---|---|---|---|---|---|---|
| 1 | 19950822 | 1256 | 85.45 | 42.38 | 197.09 | 14.06 | 99.64 |
| 2 | 19970109 | 0052 | 208.79 | 28.07 | 118.19 | 2.41 | 139.35 |
| 3 | 19970515 | 0115 | 150.20 | 74.66 | 89.82 | 53.27 | 103.76 |
| 4 | 19970715 | 0215 | 19.50 | 9.90 | 63.46 | 10.77 | 22.35 |
| 5 | 19971010 | 1557 | 64.41 | 24.90 | 62.48 | 4.70 | 27.70 |
| 6 | 19971122 | 0912 | 198.94 | 98.79 | 144.61 | 75.63 | 168.50 |
| 7 | 19980304 | 1102 | 84.05 | 41.99 | 70.94 | 41.41 | 40.19 |
| 8 | 19981018 | 1929 | 128.17 | 30.16 | 89.00 | 19.93 | 57.42 |
| 9 | 20000811 | 1849 | 106.38 | 113.98 | 90.60 | 56.96 | 154.63 |
| 10 | 20010319 | 1133 | 107.18 | 47.99 | 67.30 | 70.81 | 60.64 |
| 11 | 20010404 | 1441 | 59.08 | 211.11 | 146.90 | 123.27 | 327.35 |
| 12 | 20010421 | 1529 | 80.23 | 27.70 | 115.75 | 60.64 | 60.14 |
| 13 | 20011031 | 1347 | 64.20 | 68.53 | 210.07 | 54.84 | 301.31 |
| 14 | 20011124 | 0454 | 86.78 | 75.84 | 33.69 | 29.39 | 46.66 |
| 15 | 20020518 | 1946 | 158.76 | 160.83 | 202.78 | 240.26 | 259.96 |
| 16 | 20020801 | 0519 | 57.15 | 100.17 | 129.64 | 34.11 | 296.22 |
| 17 | 20040724 | 0531 | 140.13 | 68.46 | 168.05 | 192.04 | 267.61 |
| 18 | 20041107 | 1759 | 123.46 | 160.31 | 142.38 | 90.91 | 89.20 |
| 19 | 20050515 | 0210 | 484.94 | 298.62 | 358.26 | 469.96 | 803.67 |
| 20 | 20050612 | 0648 | 379.41 | 37.37 | 30.07 | 59.08 | 48.23 |
| 21 | 20050614 | 1756 | 253.53 | 82.10 | 78.87 | 107.21 | 213.60 |
| 22 | 20060413 | 1121 | 113.57 | 35.12 | 57.57 | 45.34 | 79.13 |
| 23 | 20071119 | 1722 | 82.14 | 34.85 | 142.13 | 49.76 | 42.87 |
| | AVERAGE | | 140.72 | 81.47 | 122.16 | 82.90 | 161.31 |

NOTE:
a Event number
b The date of event, formatted as YearMonthDay.
c The beginning time of the event, formatted as HourMinute in Table 1a and Table 1c; HourMinuteSecond in Table 1b (UT)
d Event duration (Table 1a: minute; Table 1b: second)
e The change of total magnetic field (%)
f The absolute difference of proton velocity (km/s)
g The change proton of density (%)
h The change of electron temperature (%)
i The change of proton temperature (%)
The obtained changes of local plasma and magnetic parameters are similar to the flux variations described in the second section of the paper. They have the same time ranges as the flux variations to each event. (Changes of BL: BL to upstream solar wind; changes of RE: RE to upstream solar wind; changes of shock: downstream to upstream solar wind)

**Figures**

Figure 1

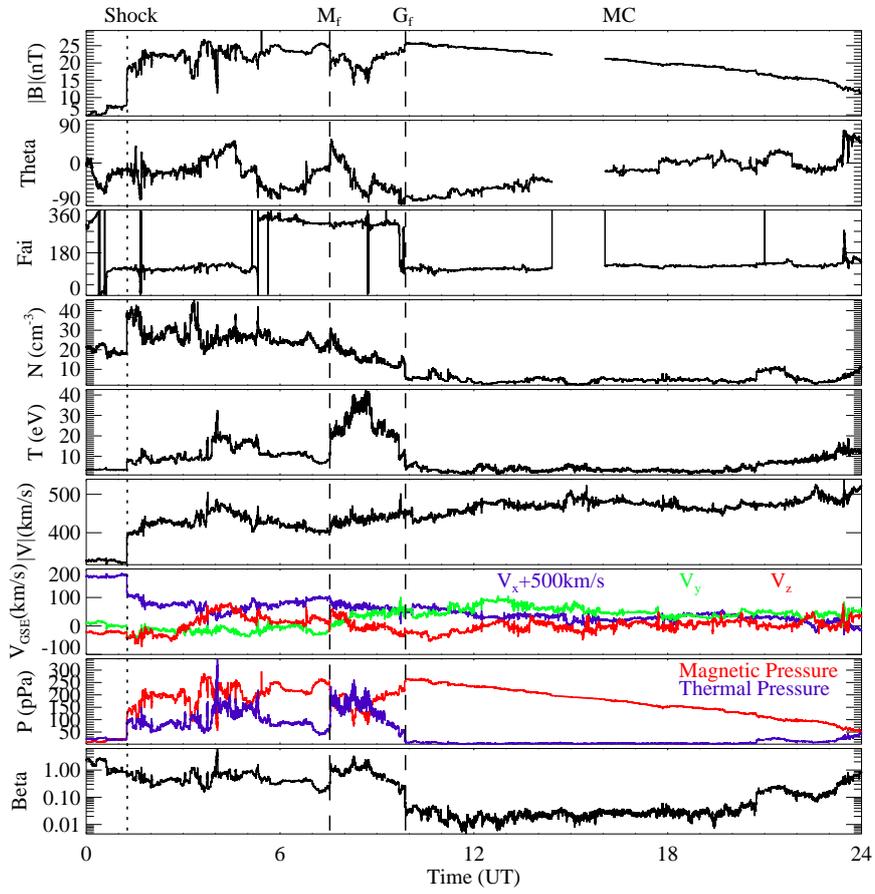

Figure 2

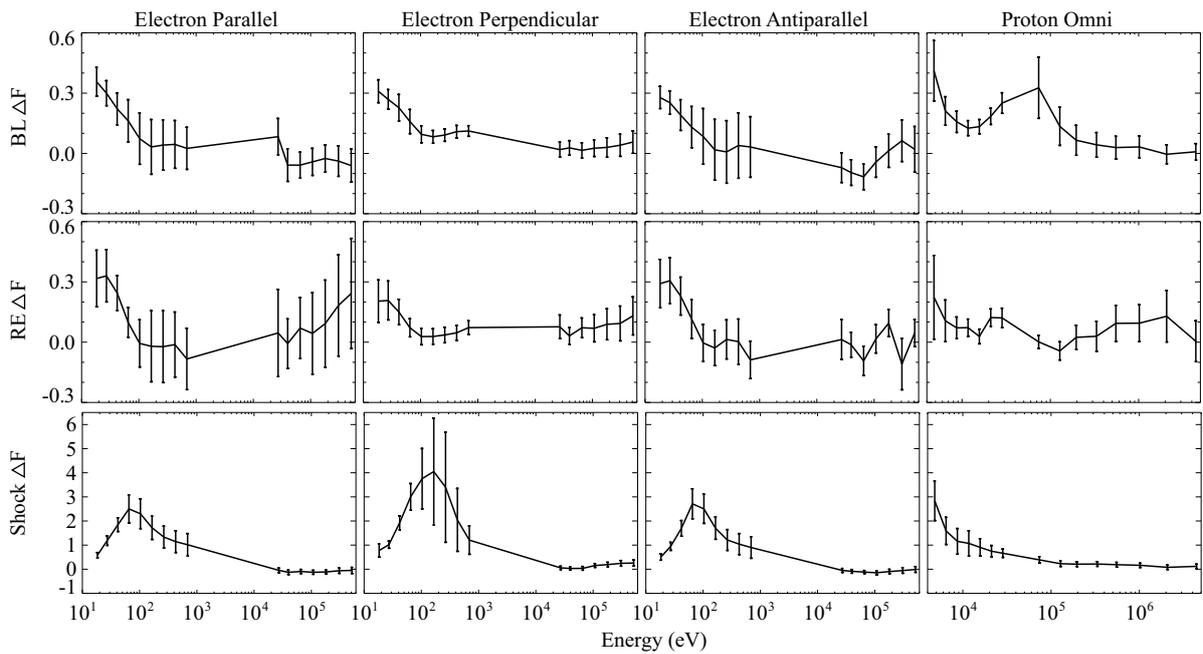

Figure 3

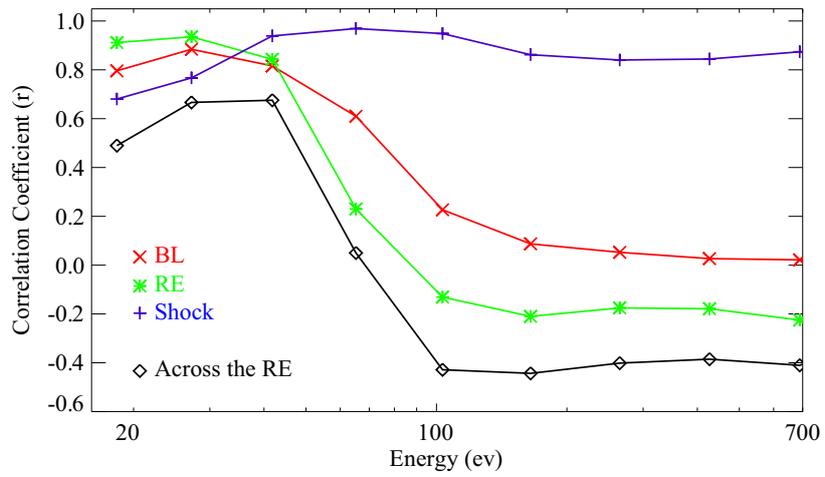

Figure 4

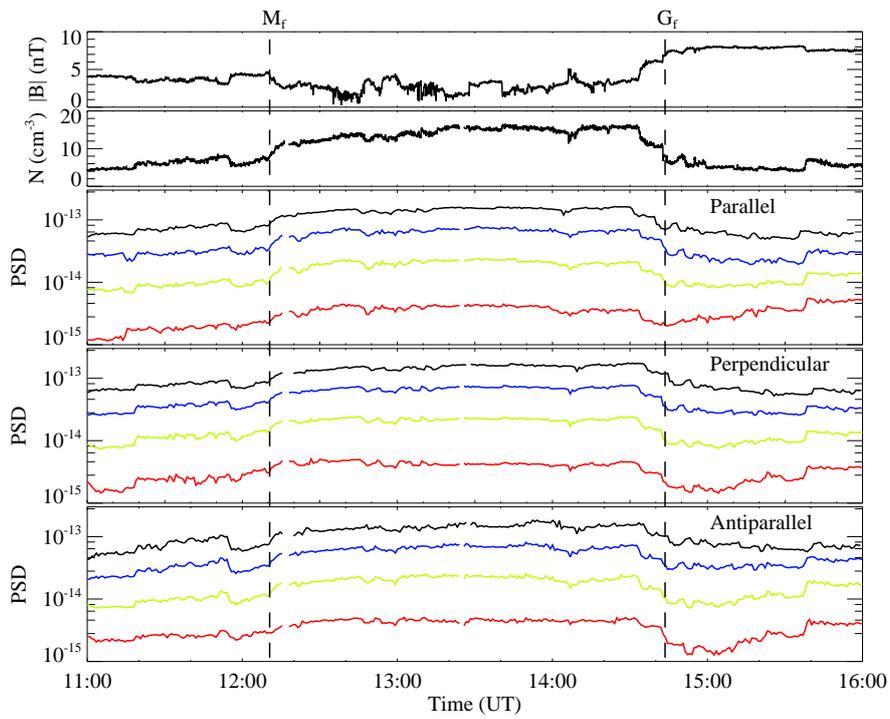

Figure 5

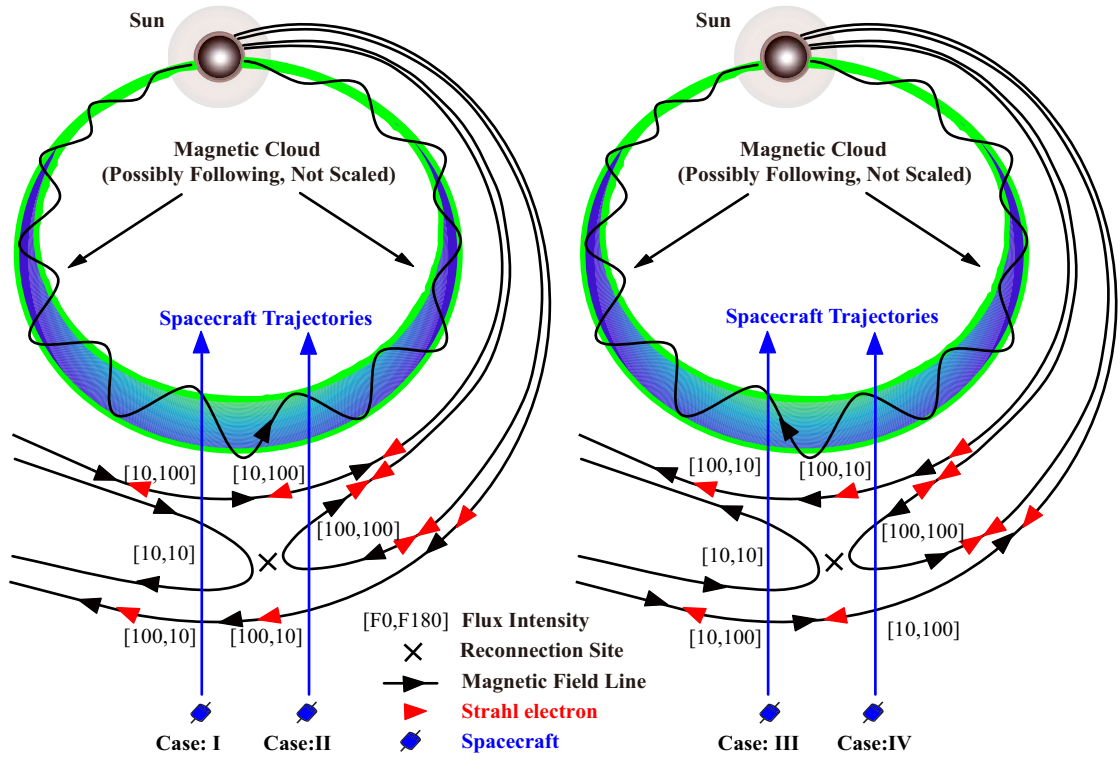